Title: Lattice Boltzmann Simulation of Non-Ideal Fluids

\documentstyle[aps,preprint]{revtex}

\title{Lattice Boltzmann Simulation of
Non-Ideal Fluids}
\author{Michael R. Swift, W. R. Osborn and J. M. Yeomans \\
	Theoretical Physics, 1 Keble Road, Oxford OX1 3NP, U.K.}

\begin{document}
\newcommand{\be}{\begin{equation}}
\newcommand{\ee}{\end{equation}}
\newcommand{\grad}{\underline \nabla}
\newcommand{\delsqu}{\nabla^{2}}
\newcommand{\ve}[1]{\underline{#1}}
\newcommand{\dpa}[2]{\frac{\partial #1}{\partial #2}}

\maketitle
\begin{abstract}
A lattice Boltzmann scheme able to model the hydrodynamics
of phase separation and two-phase flow is described.
Thermodynamic consistency is ensured by introducing a
non-ideal pressure tensor directly into the collision operator.
We also show how an external
chemical potential can be used to supplement standard boundary
conditions in order to investigate the effect of
wetting on phase separation and fluid flow in confined geometries.
The approach has the additional advantage of reducing many of the
unphysical discretisation problems common to previous lattice
Boltzmann methods.
\end{abstract}

\newpage
The hydrodynamics and kinetics of two-component fluids present a
wealth of physical problems of both fundamental and technological
importance \cite{S93}. There is much current interest in the relevance of
hydrodynamics to spinodal decomposition \cite{ACG93} and the effects
of substrates with different wetting properties on  phase
separation and domain growth \cite{T93}. In addition, the
flow properties of multicomponent systems, particularly in porous
media, have been intensively studied and are of great relevance to oil
recovery \cite{GR93}.

Conventional methods for simulating two-phase flow include numerical
integration of the Navier-Stokes equations and molecular dynamics
simulations \cite{MMBK92}. These techniques are extremely computationally
intensive and
particularly difficult to implement in random geometries. A newer
approach, the lattice Boltzmann method, has recently proved
competitive \cite{BSV92}. Here a set of distribution
functions defined on a lattice is allowed to relax to equilibrium via
a Boltzmann equation, discrete in both space and time. The correct choice of
equilibrium distribution ensures that in the long wavelength limit the
Navier-Stokes equations are recovered.

Several authors have set up lattice Boltzmann schemes for two-phase
systems. In most approaches interface
formation has been introduced phenomenologically by modifying
the Boltzmann collision operator to impose phase
separation \cite{GRZZ91}. Recent work by Shan and Chen \cite{SC93}
has attempted to relate phase
separation to microscopic interactions by redefining the equilibrium velocity
distribution so as to simulate a fluid with a non-ideal equation of state.
However, their approach leads to inconsistent
thermodynamics unless a particular equation of state is chosen.
In addition, all current schemes reach equilibrium distributions which have
unphysical velocity fluctuations within the interfacial region
\cite{GT92}.

In this letter we show for the first time that it is possible to set up
a lattice Boltzmann scheme modelling
isothermal hydrodynamics for two-phase systems. This is achieved
by introducing directly into the collision operator
the equilibrium pressure tensor for a non-ideal fluid.
The resulting phase transition is pressure driven, as pertinent
to a liquid-vapour system quenched to well below the critical
point\cite{B94}. The fluid reaches the correct thermodynamic equilibrium
as determined by the equation of state and a Maxwell construction.

We first summarise the relevant results from the Van-der-Waals
formulation of quasi-local thermodynamics for a two-component fluid
in thermodynamic equilibrium at a fixed temperature \cite{RW82}.
The free energy functional is taken to be
\be
\Psi({\ve{r}}) = \int \left\{ \frac{\kappa}{2} | \grad n({\ve{r}})|^{2}
+ \psi(n({\ve{r}})) \right\} d{\ve{r}}\, ,
\ee
where the first term gives the contribution from any density gradients
and the second describes the bulk free energy density.
The non-local pressure is defined by
\be
p(\ve{r})= n \frac{\delta \Psi}{\delta n}  - \Psi(\ve{r}) = p_{0} - \kappa
n \delsqu\!n - \frac{\kappa }{2} | \grad n|^{2} ,
\label{defp}
\ee
where $p_{0} = n \psi'(n)  - \psi(n)$ is the equation of state of the
fluid.
To obtain the full pressure tensor in a non-uniform fluid,
non-diagonal terms must be added \cite{E79}:
\be
P_{\alpha \beta}(\ve{r}) = p(\ve{r}) \, \delta_{\alpha
\beta} + \kappa \: \dpa{n}{x_{\alpha}} \! : \! \dpa{n}{x_{\beta}} .
\label{ff}
\ee
In equilibrium the components of the
pressure tensor define the surface tension in inhomogeneous
regions of the fluid.

We next describe how a lattice Boltzmann simulation can be set up to
ensure that these equilibrium conditions are met while incorporating
the dynamical behaviour pertinent to fluids, namely, order-parameter
conservation and hydrodynamic transport.
To illustrate our method, we choose to work in
two dimensions on a triangular lattice, the simplest geometry
that allows us to reproduce the Navier-Stokes equations. Let
$f_{i}(\ve{x},t)$ be a non-negative real number describing the distribution
function of the fluid density at site $\ve{x}$ at time $t$ moving in
direction $\ve{e}_{i}, i=1\ldots6$. The unit vectors $\ve{e}_{i} = \{
\cos (2 \pi (i-1)/6), \sin (2 \pi (i-1)/6) \} $ are the velocity
vectors along the links of the lattice. With each site is also
associated a function $f_{0}(\ve{x},t)$ which corresponds
to the component of the distribution with zero velocity.

The distribution functions evolve according to a Boltzmann equation
which is discrete in both space and time
\be
f_{i}(\ve{x}+\ve{e}_{i},t+1) - f_{i}(\ve{x},t)=\Omega_{i}(\ve{x},t) .
\label{aa}
\ee
The most convenient choice for $\Omega_{i}$ is a single relaxation
time form \cite{BGK54}
\be
\Omega_{i} = -\frac{1}{\tau} ( f_{i}- f_{i}^{eq}).
\label{bb}
\ee
The density $n$ and macroscopic velocity $\ve{u}$ are defined by
\begin{eqnarray}
n & = & \sum_{i} f_{i}, \\
n u_{\alpha} & = & \sum_{i} f_{i} e_{i \alpha},
\end{eqnarray}
and the equilibrium distribution, $f_{i}^{eq}$,
is chosen so as to reproduce the correct
dynamic equations for $n$ and $\ve{u}$.

For a one-component fluid with an ideal gas equation of state,
$f_{i}^{eq}$ is expanded as a power series in the local
velocity and the coefficients determined by local conservation of mass
and momentum and by the constraints of Galilean invariance and
isotropy of the pressure tensor.
In order to include the correct non-local thermodynamic properties
of a non-ideal fluid, additional non-local terms are needed in the
expansion for $f_{i}^{eq}$. We thus define
\begin{eqnarray}
f_{i}^{eq} & = & A + B e_{i \alpha} u_{\alpha} + C u^{2} + D u_{\alpha}
u_{\beta} e_{i \alpha} e_{i \beta} + F_{\alpha} e_{i \alpha} \nonumber \\
& & \hspace{1.0in} + G_{\alpha \beta} e_{i \alpha} e_{i \beta}, \label{cc} \\
f_{0}^{eq} & = & A_{0} + C_{0} u^{2}, \label {dd}
\end{eqnarray}
where the coefficients, now functions of $n$ and its derivatives,
can be determined by three macroscopic constraints.

The first two are, as
for the non-interacting case, local conservation of mass and momentum
\be
\sum_{i} f_{i}^{eq} = n ,
\label{gg}
\ee
\be
\sum_{i} f_{i}^{eq} e_{i \alpha} = n u_{\alpha}.
\label{ee}
\ee
The third constraint is that the pressure tensor takes the
form
\be
\sum_{i}f_{i}^{eq} e_{i \alpha} e_{i \beta} = P_{\alpha \beta} + n
u_{\alpha} u_{\beta} .
\label{hh}
\ee

The constraints (\ref{gg}), (\ref{ee}) and (\ref{hh}),
together with the equilibrium thermodynamic definitions (1)--(3),
are sufficient to
determine the coefficients in the expansions (\ref{cc}) and (\ref{dd})
\begin{eqnarray*}
A_{0} & = & n - 2(p_{0}-\kappa n \delsqu n) \\
A & = & (p_{0} - \kappa n \delsqu n)/3
\end{eqnarray*}

{\arraycolsep=1.1in
\( \begin{array}{cc}
B = n/3 & C_{0} = -n \\
C = -n/6 & D = 2 n /3
\end{array} \)
}
\begin{eqnarray}
F_{\alpha} & = & 0 \nonumber \\
G_{xx} = - G_{yy} & = & \frac{\kappa}{3} \left\{ \left(\dpa{n}{x}\right)^{2}
-\left(\dpa{n}{y}\right)^{2} \right\} \nonumber \\
G_{xy} & = & 2 \frac{\kappa}{3} \dpa{n}{x} \dpa{n}{y}
\end{eqnarray}
where, on the lattice, derivatives are simply expressed by
finite difference approximations.

The continuum hydrodynamic equations modelled by this dynamic scheme
can be determined by performing a Chapman-Enskog expansion on
the Boltzmann equation~(\ref{aa}).
To second order, the usual continuity and Navier-Stokes equations
result
\be
\dpa{n}{t} + \dpa{(n u_{\alpha})}{x_{\alpha}} = 0 ,
\label{pp}
\ee
\begin{eqnarray}
\dpa{n u_{\alpha}}{t} + \dpa{n u_{\beta} u_{\alpha}}{x_{\beta}}
& = & - \dpa{p_{0}}{x_{\alpha}} +
\nu \delsqu n{u_{\alpha}} +
\dpa{}{x_{\alpha}} \left( \lambda (n) \,
\grad \, . \, n {\underline u}\, \right),
\label{qq}
\end{eqnarray}
where $\nu=(2\tau-1)/8$ and
\be
\lambda (n) = (\tau-\frac{1}{2}) ( \frac{1}{2} - \dpa{p_0}{n} ).
\ee
Note that to this order, the only
difference between equation~(\ref{qq}) and the
Navier-Stokes equation for an ideal fluid is the
appearance of the
non-ideal pressure $p_{0}$.

To test the correctness and applicability of the approach described
above we performed simulations on a Van-der-Waals fluid for which
\begin{equation}
\psi = n T \ln \left( \frac{n}{1-nb} \right) -a n^2.
\end{equation}
We first consider the equilibrium configuration for a system with periodic
boundary conditions and zero net flow velocity.
The inset of Figure 1 shows
the coexistence curve as a function of temperature $T$, calculated from
equation (2) using a Maxwell construction.
The points show simulation data obtained
from lattices of size $256\times 256$, equilibrated for $10,000$ time
steps. Because the
correct
equilibrium thermodynamics is inherent in the model, the bulk phases
reached in the simulations obey
the Maxwell construction.
Note the wide range of
coexisting densities
that can be reached by the simulation before significant finite
difference errors come into play as the temperature is lowered.

Figure 1 also shows the equilibrium interface density
profiles
which are seen to depend, as expected, on $\kappa$ and the
temperature. Again the agreement
with a direct integration of the continuum thermodynamic equations
is excellent. The interfacial width can be varied,
typically between $\sim 2-30$ lattice sites.
This ensures that lattice anisotropy effects can be made unimportant.
Our mechanism for interface formation also reduces the magnitude of
the microscopic velocities in the interfacial region
by a factor of $\sim 10^3$ relative to other non-local models.
This is because in equilibrium, the effective pressure force
goes to zero and thus does not need to be balanced by momentum
transfer on the scale of the lattice discretisation.

In Figure 2 we emphasise the consistency between the mechanical
definition of surface tension
\begin{equation}
\sigma = R \Delta P,
\label{zz}
\end{equation}
where $\Delta P$ is the pressure difference between the inside and
outside of a spherical domain of radius $R$, and the thermodynamic
definition
\begin{equation}
\sigma = \kappa \int \left(\dpa{n}{z}\right)^2 dz
\label{sigma}
\end{equation}
for a flat interface. Agreement is seen as $R\rightarrow \infty$ with
the expected curvature correction to equation~(\ref{zz}) appearing for
$R \stackrel{<}{\sim} 10$ lattice units.

Far from the interface, the fluid obeys the usual Navier-Stokes equation
common to other lattice Boltzmann schemes. However, the dynamical
behaviour of the interface itself is of importance if domain growth is
to be correctly described by the model.
In Figure 3
the dispersion relation for capillary waves is displayed \cite{LL59}
giving a best fit of
$\omega \sim   k^{1.6}$. We attribute the slight discrepancy from the
expected dispersion relation $\omega^2 \sim k^3$ to curvature
corrections to Laplace's Law.
These results were
obtained by imposing a sine-wave of given wavevector on an interface
that had reached equilibrium in a $128\times128$ system and observing the
period of the subsequent oscillations for, typically, 500 timesteps.

Finally, we demonstrate how the addition of an external
chemical potential at the surfaces of a confined system
can be used to supplement the usual bounce-back
boundary conditions, allowing us to change the substrate properties and hence
study wetting.
Gradients in the chemical potential $\mu_{ex} (\ve{r})$ act as a
thermodynamic force on the fluid and can be included within the lattice
Boltzmann framework by modifying equation(11)
\be
\sum_i f_{i}^{eq} e_{i \alpha}= n u_{\alpha}
-\tau n \dpa{\mu_{ex}}{x_{\alpha}}
\ee
which, in turn, introduces a force into the collision operator,
equation(8),
\be
F_{\alpha}= - \frac{\tau n}{3} \dpa{\mu_{ex}}{x_{\alpha}}.
\ee
As an illustrative
example, by letting
$\mu_{ex}(\ve{r})$ differ from zero only at the boundary sites, the
affinity of the boundaries for each of the phases can be tuned in a simple,
physically appealing way.
Results in Figure 4 show how the fluid configuration
develops in a one-dimensional pore for a situation when the black,
dense,
fluid (a) wets and (b) does not wet the walls.
By modifying the functional form of $\mu_{ex}(\ve{r})$,
for example by introducing long range interactions,
equilibrium phases comprising of plugs, tubes and
capsules discussed by Liu et. al. \cite{LDHS90} can be reached.
In addition,
an important consequence  of this formalism is that the
fluid--boundary interface is diffuse.
This allows for
a reduction of many of the problems associated with surface
orientation
and standard bounce-back boundary conditions \cite{DPRIV}.

To summarise, we have described a lattice Boltzmann scheme, the
main new features of which are the
direct introduction of a non-ideal pressure
tensor and an external chemical potential.
This enables us to obtain an isothermal
model of phase separation which correctly describes bulk and
interfacial dynamics at low temperatures.
The method also
provides a convenient, physically
motivated way of tuning boundary conditions, giving a
new approach to situations when flow and
phase separation are affected by fluid-substrate interactions.
Moreover, unphysical velocity
oscillations at surfaces and interfaces are substantially reduced.

The simplicity of the method and the ease of implementation
suggests that our approach may be a valuable tool in the study of multi-phase
hydrodynamical systems. Furthermore, the introduction of a temperature
parameter in our mechanism for interface formation should allow
the formalism to be extended to non-isothermal situations
where heat transfer is important \cite{ACS93}.
We are currently in the process of
investigating such a model.

\vskip 0.2 truein

We are indebted to Peter Coveney, Bruce Boghosian and Tim Newman for
many stimulating discussions.

\newpage

FIGURE CAPTIONS

\begin{description}

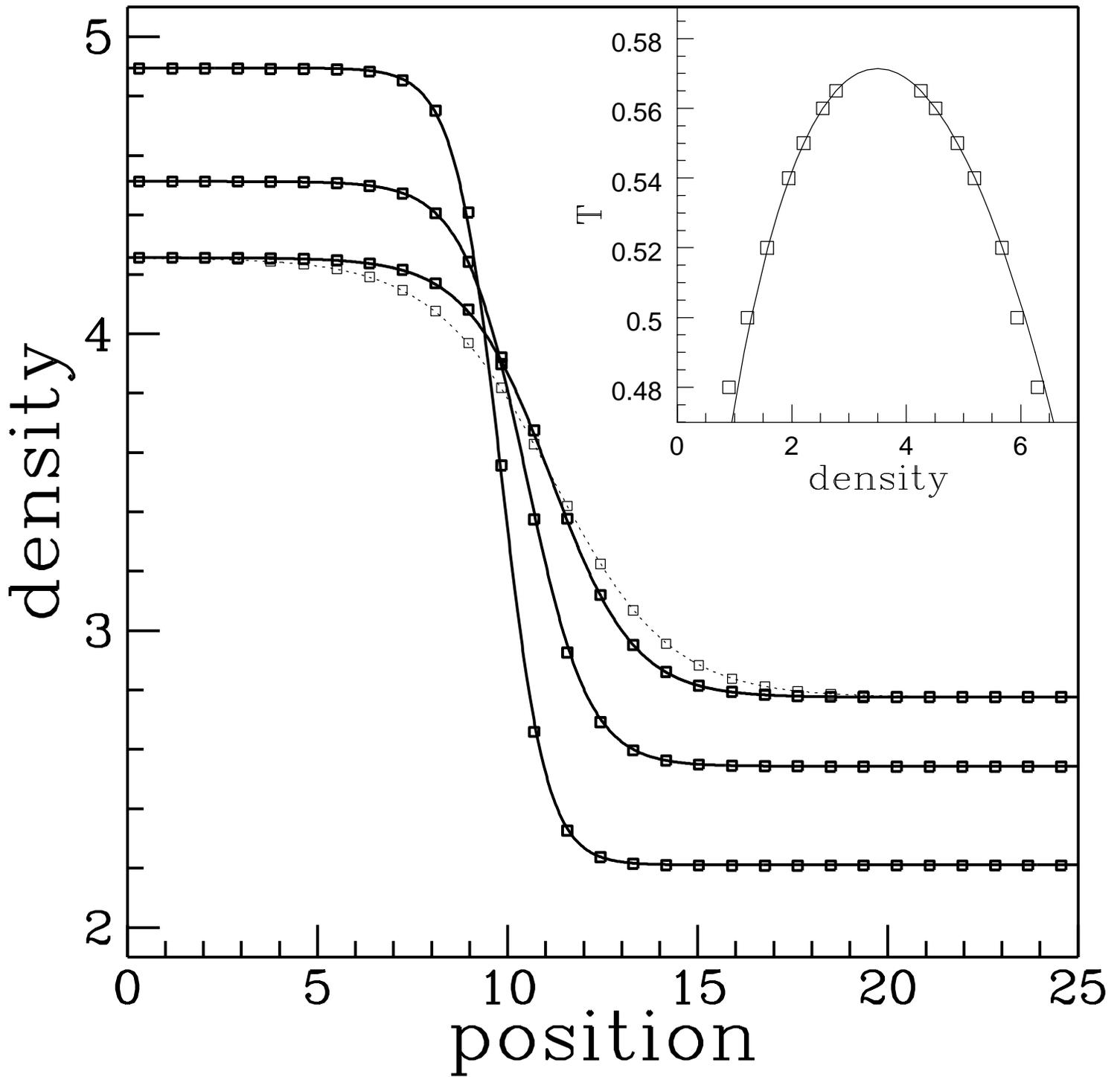
\item[Fig. 1] Equilibrium
density profiles normal to a flat interface for a Van
der Waals fluid for the three
highest values of T shown on the coexistence curve
(inset). The solid lines are numerical solutions of the
continuum thermodynamic equations while the points are from the
lattice Boltzmann simulations.
The parameter values are $a= 9/49$ and $b=2/21$, while $\kappa=0.01$ for
the bold curves and $0.02$ for the dashed curve.

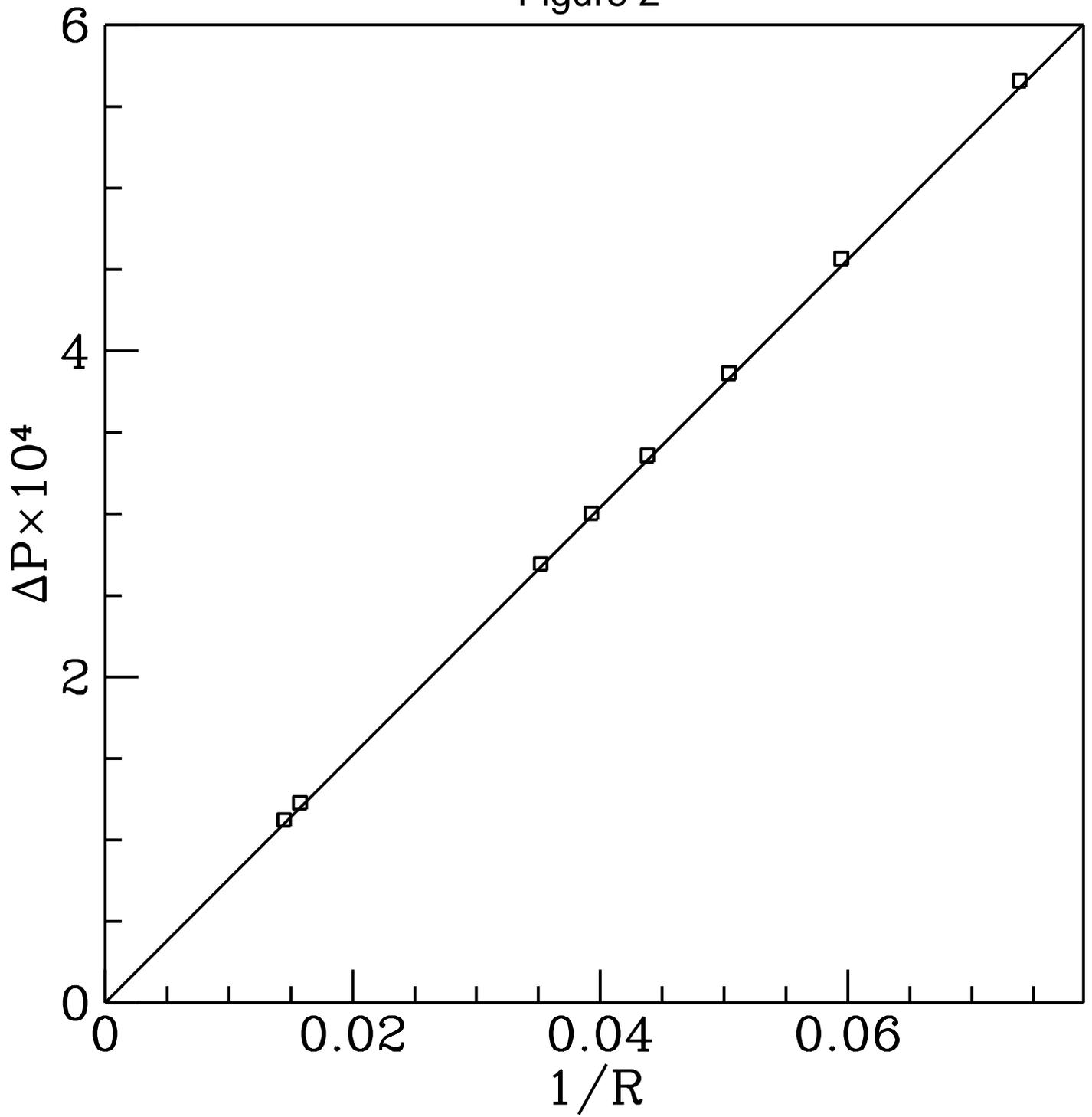
\item[Fig. 2] $\Delta P $ plotted vs. $1/R$ as a test of Laplace's Law. The
solid line has a gradient calculated using equation (\protect\ref{sigma}).

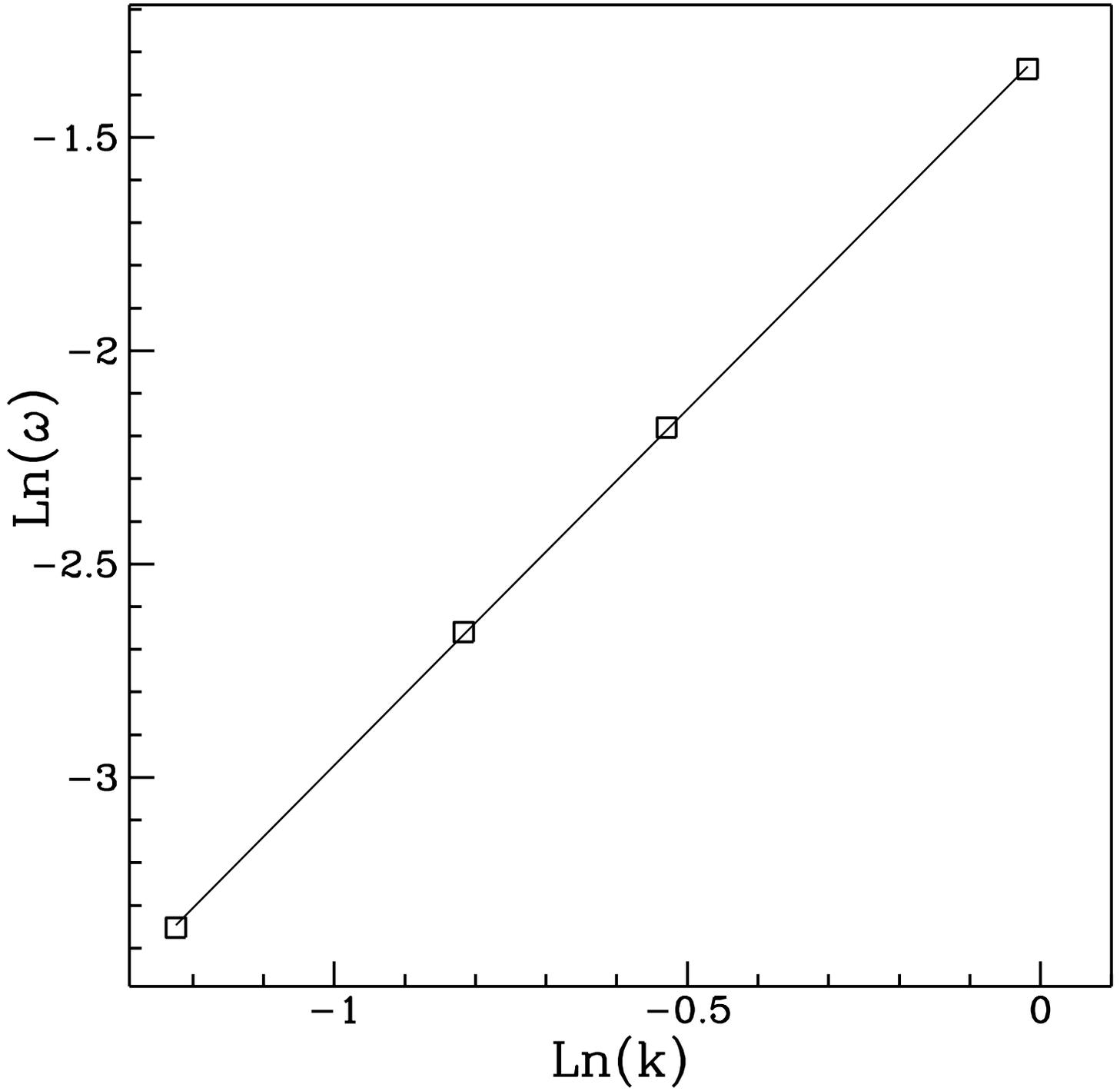
\item[Fig. 3] The dispersion curve for capillary waves on an interface,
plotted on a log-log scale. The best fit line has gradient $1.6 \pm
0.05$.

\item[Fig. 4] Time evolution (vertically downwards) of phase separation
in a narrow capillary. In (a) the dense (dark) fluid wets the surfaces
while in (b) non-wetting is illustrated. The simulations were performed
on a $128 \times 32$ lattice at $T=0.56$ with (a) $\mu_{ex}=-0.1$ and
(b) $\mu_{ex}=0.5$.

\end{description}

\centerline{     }
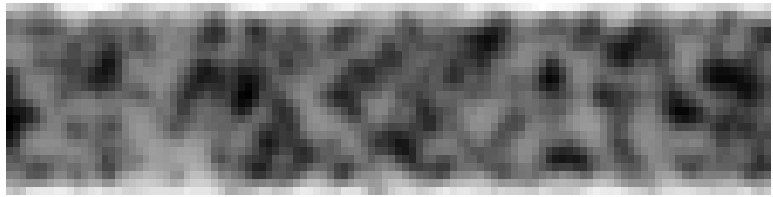
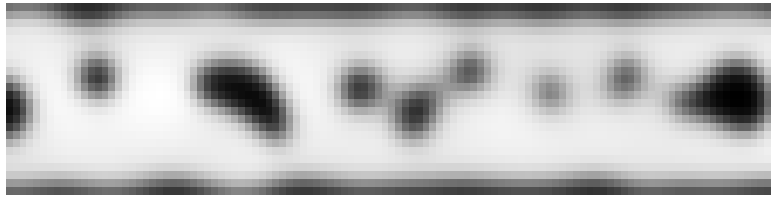
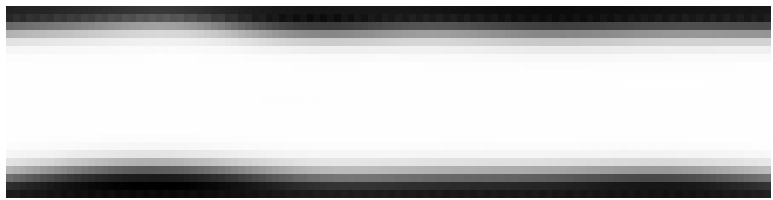
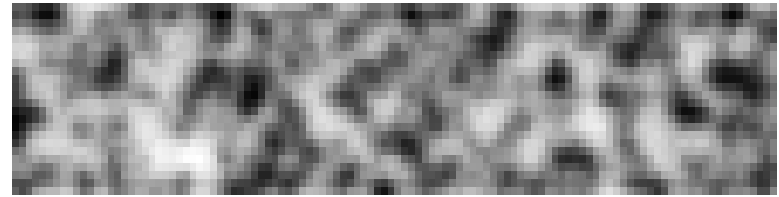
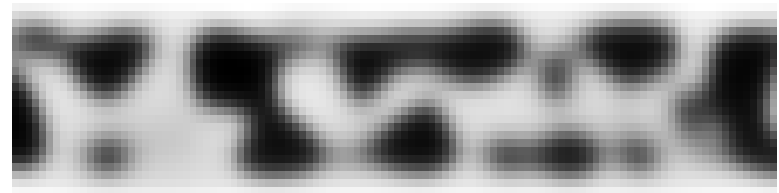
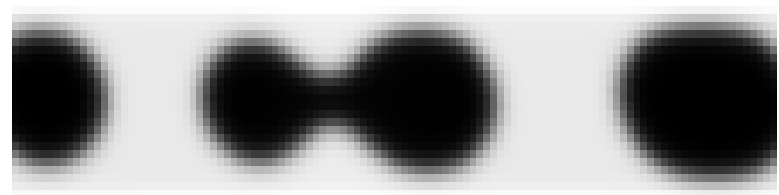
\end{document}